# The linear potential and the Dirac equation

**Walter S. Jaronski**

Department of Physics, Radford University, Radford, Virginia 24142

(wjaronsk@radford.edu)

The solution of the Dirac equation for an attractive linear potential is considered. The Lorentz nature of the potential (vector or scalar) affects the existence of bound states. For simplicity, and since it is sufficient for the goals of this study, only the ground state is considered. The case of equal vector and scalar pieces of the linear potential is emphasized because it lends itself to a simple analytic solution. This solution corresponds to a state which is strictly bound. For a linear potential with a larger component of the vector part, we find a state that is only quasi-bound, but its decay can be strongly inhibited by an effective potential barrier.

## I. INTRODUCTION

The potential included in the Dirac equation can be either the zeroth component of a Lorentz vector (like the energy) or a Lorentz scalar (like the mass). For a spherically symmetric vector potential $V(r)$ and scalar potential $S(r)$, the Dirac equation is (c = 1)

$$[\vec{\alpha} \cdot \vec{p} + \beta(m + S(r)) + V(r)]\psi(\vec{r}) = E\psi(\vec{r}) . \tag{1}$$

In this case of spherical symmetry, the spin-angular part of the solution can be written in terms of spherical spinors:

$$\psi(\vec{r}) = \begin{pmatrix} f(r)\phi_{jm}^{l} \\ ig(r)\phi_{jm}^{l'} \end{pmatrix}, \tag{2}$$

where

$$\phi_{jm}^{l} = \sum_{m_l, m_s} C(l\tfrac{1}{2}j, m_l m_s m) Y_{l m_l}(\theta, \phi) \chi_{\tfrac{1}{2} m_s}, \tag{3}$$

and where

$$l' = \begin{cases} l+1 & if\ j = l+1/2 \\ l-1 & if\ j = l-1/2 \end{cases} . \tag{4}$$

(See, e.g., Ref. 1 or Ref. 2.) Then, if we define

$$k = \begin{cases} -(j+1/2) & if\ j = l+1/2 \\ +(j+1/2) & if\ j = l-1/2 \end{cases}, \tag{5}$$

we can reduce the Dirac equation to two coupled radial equations ($\hbar = c = 1$):

$$\frac{df}{dr} + \frac{(1+k)}{r}f - (E - V + m + S)g = 0 \tag{6}$$

$$\frac{dg}{dr} + \frac{(1-k)}{r}g + (E - V - m - S)f = 0 . \tag{7}$$

The solution of these equations for a vector, attractive $1/r$ potential is well-known (the hydrogen atom).[1,2] But interest in the solution of these equations remains high because of attempts to model quarkonium states (bound states of a quark and antiquark) using the so-called "funnel" or Cornell potential

$$V = -\frac{\alpha}{r} + \lambda r . \tag{8}$$

Use of this potential in a nonrelativistic calculation with the Schrödinger equation causes no difficulties and, in fact, provides good agreement with the bound states of heavy quarkonia (if spin-dependent terms are included).[3] A potential based off of this can also be applied to good effect in semi-relativistic calculations using the spinless Salpeter equation[4]

$$\left(\sqrt{\vec{p}^2 + m_1^2} + \sqrt{\vec{p}^2 + m_2^2} + V\right)\psi = E\psi . \tag{9}$$

A difficulty arises when one attempts to use a potential of this type in the Dirac equation and treats the long-range part as a vector. It has been known for a long time that the Dirac equation with an attractive linear potential of the vector type does not yield bound states.[5] The source of the problem is the Klein paradox[6] – which shows up in this case as the possibility of the solution tunneling into free negative energy states. This behavior does not occur if the linear potential is treated as a Lorentz scalar or if the potential has a scalar component of 50% or greater.[7]

    We are not considering the more difficult case of the full two-body problem in this study. This might be addressed by the use of two-body Dirac equations from constraint dynamics.[8] But our one-body results may apply directly to the case of heavy-light quarkonia (D or B mesons, perhaps) in which the much heavier quark may be considered the source of a static potential, at least to some level of approximation. But, in addition, the nature of the Lorentz nature of the linear potential in the one-body case should shed some light on the general two-body problem.

    In this paper, we will first consider a potential which is either all scalar or all vector. We will then concentrate on the case of a linear potential which is half vector and half scalar. Although, as already stated, we are restricted to the one-body case, it is interesting to note that this mix is also used in the two-body treatment of Ref. 8.

## II. VECTOR AND SCALAR CASES

For our purposes it will be sufficient to consider the ground state problem, for which $j = ½$, $l = 0$, and $k = -1$. With that choice and taking a pure scalar potential, we have

$$\frac{df}{dr} - (E + m + S) = 0 \tag{10}$$

$$\frac{dg}{dr} + 2\frac{g}{r} + (E - m - S) = 0 . \tag{11}$$

Defining

$$f = \frac{u}{r}, \quad g = \frac{v}{r}, \tag{12}$$

gives

$$\frac{du}{dr} - \frac{u}{r} - (E + m + S)v = 0 \tag{13}$$

$$\frac{dv}{dr} + \frac{v}{r} + (E - m - S)u = 0 . \tag{14}$$

We differentiate Eq. (13):

$$\frac{d^2u}{dr^2} - \frac{1}{r}\frac{du}{dr} + \frac{u}{r^2} - (E + m + S)\frac{dv}{dr} - \frac{dS}{dr}v = 0 . \tag{15}$$

Substituting for $dv/dr$ from Eq. (14) and $v$ from Eq. (13) yields

$$(E + m + S)\frac{d^2u}{dr^2} - \frac{dS}{dr}\frac{du}{dr} + \frac{dS}{dr}\frac{u}{r} + (E + m + S)^2(E - m - S)u = 0 . \tag{16}$$

For an attractive linear potential, $S = \lambda r$ ($\lambda > 0$), this is

$$(E + m + \lambda r)\frac{d^2u}{dr^2} - \lambda\frac{du}{dr} + \lambda\frac{u}{r} + (E + m + \lambda r)^2(E - m - \lambda r)u = 0 . \tag{17}$$

Now consider the limit $r \to \infty$. In this limit, we have

$$\lambda r \frac{d^2 u}{dr^2} - \lambda^3 r^3 u = 0 \tag{18}$$

and it is easily verified that an asymptotic solution of this equation is

$$u = A e^{-\lambda r^2/2}, \tag{19}$$

indicative of a bound state. (Of course, the positive argument of the exponential also solves the equation, but we reject this exponentially growing solution for the obvious physical reason of finiteness at large $r$.) So, it appears that bound states exist in this case; and, in fact, for certain choices of the parameters, analytic bound state solution can be found.[9]

For a vector potential, our equations for the reduced radial wave functions become

$$\frac{du}{dr} - \frac{u}{r} - (E + m - \lambda r)v = 0 \tag{20}$$

$$\frac{dv}{dr} + \frac{v}{r} + (E - m - \lambda r)u = 0. \tag{21}$$

Proceeding in a similar fashion as above, this gives

$$(E + m - \lambda r)\frac{d^2 u}{dr^2} - \lambda \frac{du}{dr} - \lambda \frac{u}{r} + (E + m - \lambda r)^2 (E - m - \lambda r)u = 0. \tag{22}$$

In the large-$r$ limit, this becomes

$$-\lambda r \frac{d^2 u}{dr^2} - \lambda^3 r^3 u = 0, \tag{23}$$

which is solved at large $r$ by

$$u = A e^{\pm i \lambda r^2/2}, \tag{24}$$

indicative of an unbound state. We conclude that any "bound" state in this case is, at best, quasi-bound and can decay. The classically allowed region for positive energy states is $r < r_1$, where

$$m + \lambda r_1 = E \Rightarrow r_1 = \frac{E - m}{\lambda}. \tag{25}$$

In this region, the kinetic energy is positive and the wave function should have a typical bound-state form. For $r > r_1$, we expect the wave function to be decaying. However, if the potential is able to lift the negative-energy continuum to positive energies, then at a large enough value of $r$, the wave function will contain a component corresponding to free negative-mass states. This will occur when the rising line of the top of the negative-mass continuum, $-m + \lambda r$, reaches the value of the energy, i.e., at $r = r_2$, where

$$-m + \lambda r_2 = E \Rightarrow r_2 = \frac{E + m}{\lambda}. \tag{26}$$

We can obtain some understanding of this by considering our equation for $u$. Make the change of variable

$$x = E + m - \lambda r. \tag{27}$$

Then the equation becomes

$$x \frac{d^2 u}{dx^2} + \frac{du}{dx} - \frac{u}{E + m - x} + \frac{x^2}{a^2}(x - 2m)u = 0, \tag{28}$$

and we note that when $r = r_2$, $x = 0$ and when $r = r_1$, $x = 2m$. The region $r > r_2$ (the region of free "lifted" negative-mass states) is $x < 0$; the region $r < r_1$ (the region of the quasi-bound state) is $x > 2m$. The interval between $r_1$ and $r_2$ is classically forbidden but, quantum mechanically, "tunneling" can occur between the quasi-bound region and the free negative-mass states; the classically forbidden region corresponds to the interval $0 < x < 2m$. Near $x = 0$, we have the approximate equation

$$x \frac{d^2 u}{dx^2} + \frac{du}{dx} - \frac{u}{E + m} = 0. \tag{29}$$

The solution of this equation is

$$u = \begin{cases} AJ_o\left(2\sqrt{-\frac{x}{E + m}}\right), & x < 0 \\ AI_o\left(2\sqrt{\frac{x}{E + m}}\right), & x > 0 \end{cases} \tag{30}$$

where $J_0$ is the Bessel function of order zero, and $I_0$ is the modified Bessel function (hyperbolic Bessel function) of the first kind of order zero. The functions of the second kind are not included, since they are divergent at $x = 0$. The function is continuous at $x = 0$ since $J_0(0) = I_0(0) = 1$. Note that the function is oscillatory for $x < 0$ ($r > r_2$), and decaying to its value at $x = 0$, for $x > 0$ ($r < r_2$). This is exactly the behavior we would expect for a state in the classically forbidden region joining to a free state at the edge of the tunneling region.

Near $x = 2m$ ($r = r_1$), the wave function satisfies the approximate equation

$$x\frac{d^2u}{dx^2} + \frac{du}{dx} - \frac{u}{E-m} = 0 \,. \tag{31}$$

The solution of this is

$$u = BI_0\left(2\sqrt{\frac{x}{E-m}}\right) + CK_0\left(2\sqrt{\frac{x}{E-m}}\right), \tag{32}$$

where $K_0$ is the modified Bessel function of the second kind of order zero, and we note that $E > m$ for our quasi-bound state. This has a finite value at $x = 2m$, and the constants $B$ and $C$ should be chosen to give a function decreasing from $r < r_1$ to $r > r_1$. Proceeding in this direction, the $I_0$ function will decrease and the $K_0$ function will increase. However, for most common values of the parameters, $I_0$ will be much larger than $K_0$ in the neighborhood of $r_1$ and a dominant mixture of $I_0$ in Eq. (32) will give the expected behavior. The exact ground-state wave function, starting at $u(0) = 0$ and reaching a maximum in the range $0 < r < r_1$, will join to this approximate function at $r = r_1$ and then also provide the bridge between the approximate solution at $r = r_1$ and $r = r_2$. Numerical solutions of Eqs. (20) and (21) confirm this behavior.

The quasi-bound state is separated from the continuum by the large "tunneling" region, $r_2 - r_1 = 2m/\lambda$. A related discussion of this situation is presented in the case of one spatial dimension in Ref. 10. An approximate value of the lifetime for our quasi-bound state may be obtained by using the old Gamow/Condon-Gurney argument for alpha decay.[11] Application of that technique in this case is not intended to give a rigorous result, but it merely provides a way to quantify the relative amount of tunneling (ultimately we will use it for different mixtures of the vector and scalar components in the potential). We will use the Gamow expression for the tunneling probability with the simple change of replacing the modulus of the momentum by its relativistic expression:

$$|p| = \sqrt{m^2 - (E - V(r))^2} \,. \tag{33}$$

The lifetime is given by

$$\tau = \tau_0 e^{2\gamma} \tag{34}$$

where

$$\gamma = \int_{r_1}^{r_2} \sqrt{m^2 - (E - \lambda r)^2}\, dr \,. \tag{35}$$

and where $\tau_0$ is a time characteristic of the scale of the problem. In this picture, $\tau_0$ is the time between attempts to penetrate the barrier; so $\tau_0 \sim r_1/<v>$, and, for example, if $r_1$ is of the order of a fermi (femtometer) and the average speed $<v>$ is a significant fraction of $c$, then $\tau_0 \sim 10^{-24}$ s.

This procedure should suffice since we are only interested in an order-of-magnitude estimate. A more detailed treatment requires a discussion of the WKB approximation for the Dirac equation.[12] The integral of Eq. (35) is elementary. One obtains

$$\gamma = \frac{\pi m^2}{2\lambda}. \qquad (36)$$

This can vary widely in models. For $\tau \approx 10^3 \tau_0$, we require $\gamma \approx 3.5$. One would be well-advised to consider this parameter carefully before using a 100% vector linear potential in a model.

We note, in passing, an interesting result obtained by Sauter in 1931.[13] He considered the Dirac equation in one dimension with the Klein step potential replaced by a linear ramp: $V(x) = vx$, between regions of constant potential, with $V = 0$ for $x < 0$ and $V(x) = vL$, for $x > L$. He finds a transmission probability to negative-energy states for $v$ in the range $2m/L < v < m^2$ of order $e^{-\pi m^2/v}$, i.e., a probability essentially governed by the same quantity as that of Eq. (36).

## III. SOLUTION OF DIRAC EQUATION FOR A LINEAR POTENTIAL WITH EQUAL PARTS VECTOR AND SCALAR

If the potential has equal vector and scalar pieces, the equations (6) and (7) have a particularly simple form, which may allow elementary analytic solutions.[14] We take

$$V = S = \frac{1}{2}\lambda r \qquad (37)$$

($\lambda > 0$) in Eqs. (6) and (7) to get

$$\frac{df}{dr} + \frac{(1+k)}{r}f - (E+m)g = 0 \qquad (38)$$

$$\frac{dg}{dr} + \frac{(1-k)}{r}g + (E - m - \lambda r)f = 0. \qquad (39)$$

For simplicity, we limit ourselves again to the lowest energy state: $l = 0, j = ½$, for which $k = -1$:

$$\frac{df}{dr} - (E+m)g = 0 \qquad (40)$$

$$\frac{dg}{dr} + \frac{2}{r}g + (E - m - \lambda r)f = 0. \tag{41}$$

Introducing, as before, the modified radial wave functions $u$ and $v$:

$$f = \frac{u}{r}, \quad g = \frac{v}{r}, \tag{42}$$

gives

$$\frac{du}{dr} - \frac{u}{r} - (m + E)v = 0 \tag{43}$$

$$\frac{dv}{dr} + \frac{v}{r} + (E - m - \lambda r)u = 0. \tag{44}$$

We now rewrite Eq. (43) as an equation for $v$, differentiate this, and substitute $v$ and its derivative into Eq. (44) to yield

$$\frac{d^2 u}{dr^2} - (m + E)(m - E + \lambda r)u = 0, \tag{45}$$

or, with $m^2 - E^2 \equiv q^2$ ($< 0$),

$$\frac{d^2 u}{dr^2} - q^2 u - \lambda(m + E)ru = 0. \tag{46}$$

We now make the change of variables

$$\xi = [\lambda(m + E)]^{1/3}\left(r + \frac{q^2}{\lambda(m + E)}\right), \tag{47}$$

to obtain

$$\frac{d^2 u}{d\xi^2} - \xi u = 0, \tag{48}$$

which is Airy's equation. Taking the solution which does not blow up at large $r$,

$$u = c_1 Ai(\xi) = c_1 Ai\left([\lambda(m + E)]^{1/3}(r + q^2/\lambda(m + E))\right), \tag{49}$$

where $c_1$ is a constant. We note that the argument of the Airy function is negative for $r < r_1$, where

$$r_1 + \frac{q^2}{\lambda(m + E)} = r_1 + \frac{m^2 - E^2}{\lambda(m + E)} = 0 \Rightarrow r_1 = \frac{E - m}{\lambda}, \tag{50}$$

and positive beyond this point. This value of $r$ is the classical turning point ($m + V = m + \lambda r = E$). The Airy function is oscillatory for negative values of its argument and decreases rapidly for positive values. Therefore, this has all the characteristics of a bound state. The eigenenergy is determined by the requirement that $u = 0$ at $r = 0$ (recall that $u = rf$, and $f$ should be regular at the origin):

$$Ai([\lambda(m + E)]^{1/3} q^2/\lambda(m + E)) = 0. \tag{51}$$

So we must have

$$\frac{q^2}{[\lambda(m + E)]^{2/3}} = \frac{m^2 - E^2}{[\lambda(m + E)]^{2/3}} = \beta_i, \tag{52}$$

where $\beta_i$ is one of the zeros of the Airy function. We are looking for the ground state and so we take the first zero:

$$\frac{m^2 - E^2}{[\lambda(m + E)]^{2/3}} = -2.338, \tag{53}$$

or

$$\frac{E^2 - m^2}{[\lambda(m + E)]^{2/3}} = 2.338. \tag{54}$$

Note that in these units, $\lambda$ has dimensions of (energy)$^2$. As a simple numerical example, if $\lambda = 0.2$ GeV$^2$ (within the range of popular choices for quarkonium models) and $m = 1$ GeV, then $E = 1.5828$ GeV.

We should also check the behavior of the lower component of the wave function at the origin. We have

$$v(r) = \frac{1}{E + m}\left(\frac{du}{dr} - \frac{u}{r}\right), \tag{55}$$

and we might be concerned by a possible singularity at the origin. But, for a function whose derivative is defined at $r_0$,

$$\frac{du}{dr} \rightarrow \frac{u(r) - u(r_0)}{r - r_0} \tag{56}$$

as $r \rightarrow r_o$. Thus, since $u(r)$ has a zero at $r = 0$, $du/dr \rightarrow u/r$ as $r \rightarrow 0$ and $v(r)$ vanishes at the origin, as it should. The general behavior of $v$ is also similar to that of $u$ – oscillatory for $r < r_1$ and rapidly decreasing beyond that point.

## IV. ARBITRARY VECTOR-SCALAR MIX

In the case of equal vector and scalar parts of the linear potential, the negative energy states are not lifted by the potential. The top of the negative-energy continuum is shifted to $-m + \lambda r/2$ by the vector potential and to $-m - \lambda r/2$ by the scalar potential and so remains unchanged. The negative-energy continuum then has no effect on the bound state of positive energy.

If the potential is a fraction $s$ scalar, that is, if

$$S(r) = s\lambda r, \qquad V(r) = (1-s)\lambda r, \tag{57}$$

with $0 \le s \le 1$, then the top of the negative energy continuum is shifted up by $(1-s)\lambda r$ by the vector and down by $s\lambda r$ by the scalar. The net shift up is $(1-s)\lambda r - s\lambda r = (1-2s)\lambda r$. If the potential is predominantly scalar ($s > \frac{1}{2}$), the shift is down and the negative-energy states have no effect upon the bound state. But if the state is predominantly vector ($s < \frac{1}{2}$), the top of the negative-energy continuum is rising and will intersect the energy line when

$$-m + (1-2s)\lambda r = E \Rightarrow r = \frac{E+m}{(1-2s)\lambda} \equiv r_3. \tag{58}$$

There is then the possibility of tunneling to these lifted negative-energy states for $r > r_3$. With our simple Gamow treatment, the quantity $\gamma$ will now be

$$\gamma = \int_{r_1}^{r_2} \sqrt{m^2 - (E-\lambda r)^2}\, dr + \int_{r_2}^{r_3} \sqrt{(E-\lambda r)^2 - m^2}\, dr, \tag{59}$$

where $r_2$ is as before (Eq. (26)), and where the change in order inside the square root in the second term is required since for $r > r_2$, $\lambda r - E > m$. Therefore,

$$\gamma = \frac{\pi m^2}{2\lambda} + \int_{r_2}^{r_3} \sqrt{(\lambda r - E)^2 - m^2}\, dr \tag{60}$$

and, although the integral depends on $E$, it is clearly positive since $r_3 > r_3$. In fact, $r_3/r_2 = 1/(1-2s)$ and $r_3$ can be significantly larger than $r_2$ for $s$ close to 0.5 (but less than 0.5 – the case $s \ge 0.5$ corresponds to a true bound state, as discussed above). So for a particular choice of parameters we can have a large value of $\gamma$ and, *a fortiori*, a large value of $e^{2\gamma}$, and therefore have negligible tunneling even if the state is predominantly vector as long as it has a significant scalar fraction. We admit that our technique for judging the boundedness of the state is only approximate. In a more detailed treatment, the degree of mixing (tunneling) may depend on the behavior of the wave function in the forbidden region. If the decay has a significant "tail" in this region, the mixing may be stronger than our analysis predicts. This might occur for excited states, for example, and then the state may strongly deviate more strongly from a bound state. Nevertheless, we believe that our analysis provides a valuable approach to the boundedness issue and indicates that in many cases a vector component to the potential may have a negligible effect on the existence of a strongly quasi-bound state. This is an important consideration for quarkonium model builders. Although the choice of a confining potential which is at least 50% scalar avoids the binding problem, the spin-dependent parts of the potential (based on the standard reduction of the Bethe-Salpeter kernel[15]) depend on the vector-scalar mixture and a large vector component might be desired to explain observed spin splittings of states of different orbital and spin angular momenta. Therefore, it is significant that a vector piece to the linear potential, which is larger than 50%, might be acceptable in a specific model.

---